\documentclass[letterpaper, 10 pt, conference]{IEEEtran}

\usepackage{tabularx}

\usepackage{algpseudocode}
\usepackage{algorithm}
\usepackage{algorithmicx}

\usepackage{subfigure}
\usepackage{lipsum} 
\usepackage{arydshln}
\usepackage[dvips]{color}
\usepackage{comment}
\usepackage{todonotes}
\usepackage{epsf}
\usepackage{epsfig}
\usepackage{times}
\usepackage{epsfig}
\usepackage{graphicx}
\usepackage{bbold}
\usepackage{mathtools}
\usepackage{mathrsfs}
\usepackage{amssymb}
\usepackage{pdfpages}
\usepackage{epstopdf}
\newfloat{algorithm}{t}{lop}

\usepackage{amsmath}
\usepackage{dsfont}
\usepackage{lettrine} 

\usepackage{amsmath,epsfig,amssymb,algorithm,algpseudocode,amsthm,cite,url}
\usepackage{caption}

\allowdisplaybreaks
\usepackage{csquotes}

\usepackage{verbatim}
\usepackage[english]{babel}
\usepackage{amsmath,amssymb}

\usepackage{multirow}
\usepackage{rotating}

\usepackage[justification=centering]{caption}
\usepackage{verbatim}

\begin{document}
	%
	 \title{Machine Learning for Wireless Connectivity and Security of Cellular-Connected UAVs}
	\IEEEoverridecommandlockouts
	\author{\IEEEauthorblockN{Ursula Challita\IEEEauthorrefmark{1}\thanks{\IEEEauthorrefmark{1}This work was done when this author was at The University of Edinburgh.}, Aidin Ferdowsi\IEEEauthorrefmark{2}, Mingzhe Chen\IEEEauthorrefmark{3}, and Walid Saad\IEEEauthorrefmark{2}\\}
			\IEEEauthorblockA{\IEEEauthorrefmark{1}
			Ericsson Research, Stockholm, Sweden, Email: {ursula.challita@ericsson.com}\\}
		\IEEEauthorblockA{\IEEEauthorrefmark{2}
			Wireless@VT, Bradley Department of Electrical and Computer Engineering, \\ Virginia Tech, Blacksburg, VA, USA,
			Emails: \{aidin,walids\}@vt.edu\\}
		\IEEEauthorblockA{\IEEEauthorrefmark{3}
			Beijing Laboratory of Advanced Information Network, Beijing University of Posts and Telecommunications, \\Beijing, China 100876, Email: {chenmingzhe@bupt.edu.cn}\\}
\vspace{-1cm}
		\thanks{This work was supported by the Army Research Office (ARO) under Grant W911NF-17-1-0593 and, in part, by the U.S. National Science Foundation under Grants OAC-1541105 and IIS-1633363. The views and conclusions contained in this document are those of the authors and should not be interpreted as representing the official policies, either expressed or implied, of ARO or the U.S. Government. The U.S. Government is authorized to reproduce and distribute reprints for Government purposes notwithstanding any copyright notation herein.}
	}
	\maketitle
	
\IEEEpeerreviewmaketitle
	
\begin{abstract}
Cellular-connected unmanned aerial vehicles (UAVs) will inevitably be integrated into future cellular networks as new aerial mobile users. Providing cellular connectivity to UAVs will enable a myriad of applications ranging from online video streaming to medical delivery. However, to enable a reliable wireless connectivity for the UAVs as well as a secure operation, various challenges need to be addressed such as interference management, mobility management and handover, cyber-physical attacks, and authentication. In this paper, the goal is to expose the wireless and security challenges that arise in the context of UAV-based delivery systems, UAV-based real-time multimedia streaming, and UAV-enabled intelligent transportation systems. To address such challenges, artificial neural network (ANN) based solution schemes are introduced. The introduced approaches enable the UAVs to adaptively exploit the wireless system resources while guaranteeing a secure operation, in real-time. Preliminary simulation results show the benefits of the introduced solutions for each of the aforementioned cellular-connected UAV application use case.
\end{abstract}

\section{Introduction}\label{Intro}
\begin{figure}[t!]
  \begin{center}
  \centering
   \includegraphics[width=9cm]{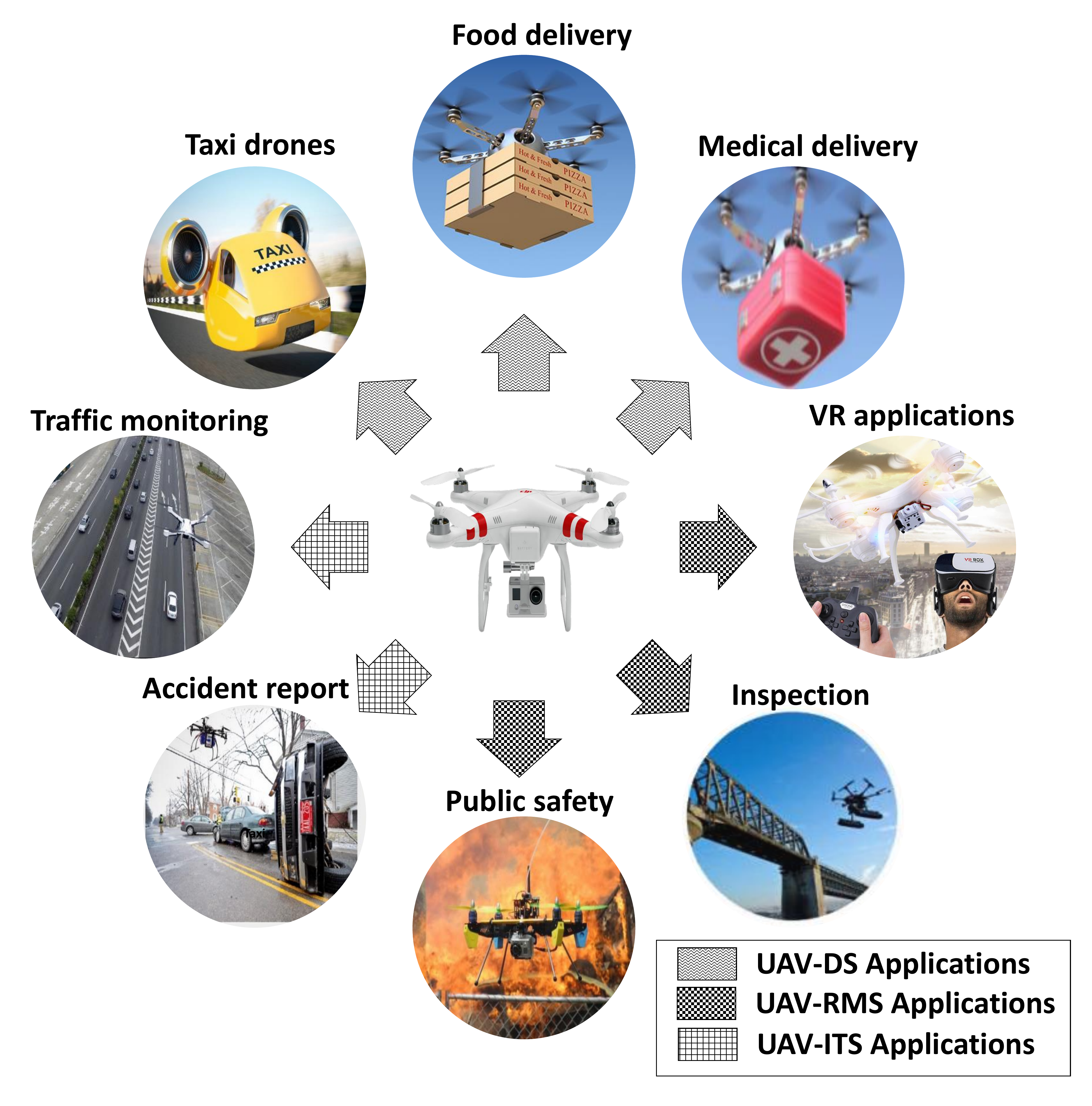}
   \caption{Cellular-connected UAVs applications in UAV-based delivery systems, UAV-based real-time multimedia streaming networks, and UAV-enabled intelligent transportation systems.}\label{UAV_applications}
  \end{center}
  \vspace{-0.6cm}
\end{figure}

Unmanned aerial vehicles (UAVs) will be ubiquitous and will play a vital role in various sectors ranging from medical and agricultural to surveillance and public safety. Providing connectivity to UAVs is crucial for data collection and dissemination in such applications. Unlike current wireless UAV connectivity that relies on short-range communication technologies (e.g., WiFi, Bluetooth), cellular connectivity allows beyond line-of-sight control, low latency, real time communication, robust security, and ubiquitous coverage. In essence, cellular-connected UAVs will lead to many new application use cases which we classify into three primary categories: UAV-based delivery systems (UAV-DSs), UAV-based real-time multimedia streaming (UAV-RMS) networks, and UAV-enabled intelligent transportation systems (UAV-ITSs), as shown in Figure~\ref{UAV_applications}.

However, to reap the benefits of cellular-connected UAVs for UAV-DSs, UAV-RMS, and UAV-ITSs use cases, various unique communication and security challenges for each of such applications need to be addressed. For instance, efficient handover and online path planning are more crucial for UAV-DSs applications while cooperative multi-UAV data transmission and secured consensus of UAV swarms are unique for UAV-ITSs. In this scope, artificial intelligence (AI) based solution schemes are regarded as a powerful tool for addressing the aforementioned challenges\footnote{For more information, technical details related to the proposed AI techniques can be found at~\cite{tutorial_ML}}. It is worthwhile noting that such challenges can also be addressed at different levels such as the PHY layer and 3D coverage enhancement\footnote{Some existing surveys already discuss some of these issues~\cite{sebastien, mozaffari_survey}.}. In this regard, AI-based solution schemes assist in addressing the aforementioned challenges while yielding new improvements in the design of the network. Although many approaches exist for addressing the aforementioned challenges, we focus on machine learning solutions\footnote{The proposed machine learning techniques are mainly divided into two phases, the training phase followed by the testing phase. Therefore, although the training phase requires some heavy computation, it does not have any impact on the behavior of the UAVs during the testing phase, which refers to the actual execution time.} due to their inherent ability for predicting future network states thus allowing the UAVs to adapt to the dynamics of the network in an online manner. In particular, machine learning techniques allow the UAVs to generalize their observations to unseen network states and can scale to large-sized networks which therefore makes them suitable for UAV applications. Moreover, for such UAV-based applications, energy efficiency and computation capability is a key design constraint. Consequently, the main scope of this work is to highlight the advantages that AI brings for cellular-connected UAVs, under various constraints.



In this regard, current existing literature study the changes in the radio environment of cellular-connected UAVs with altitude and analyze the corresponding implications on mobility performance~\cite{sebastien}. Moreover, in~\cite{mozaffari_survey}, the authors provide an overview on the opportunities and challenges for the use of UAVs for wireless communication applications; however, the primary focus is on their application as base stations (BSs). The authors in~\cite{ismail} propose a trajectory optimization scheme for cellular-connected UAVs while guaranteeing cellular connectivity. Although the works in~\cite{mozaffari_survey} and~\cite{ismail} discuss cellular-connected UAVs, they do not focus on the specifics of UAV-DS, UAV-RMS, and UAV-ITS applications, nor do they address AI or security challenges. Therefore, despite being interesting, none of the existing works propose and evaluate AI-based solutions for addressing both wireless and security challenges that arise in the context of cellular-connected UAVs. In essence, the state-of-the-art does not study the potential of AI as a means for addressing the challenges of integrating cellular-connected UAVs across various applications.

\begin{figure}[t!]
  \begin{center}
  \centering
   \includegraphics[width=7cm]{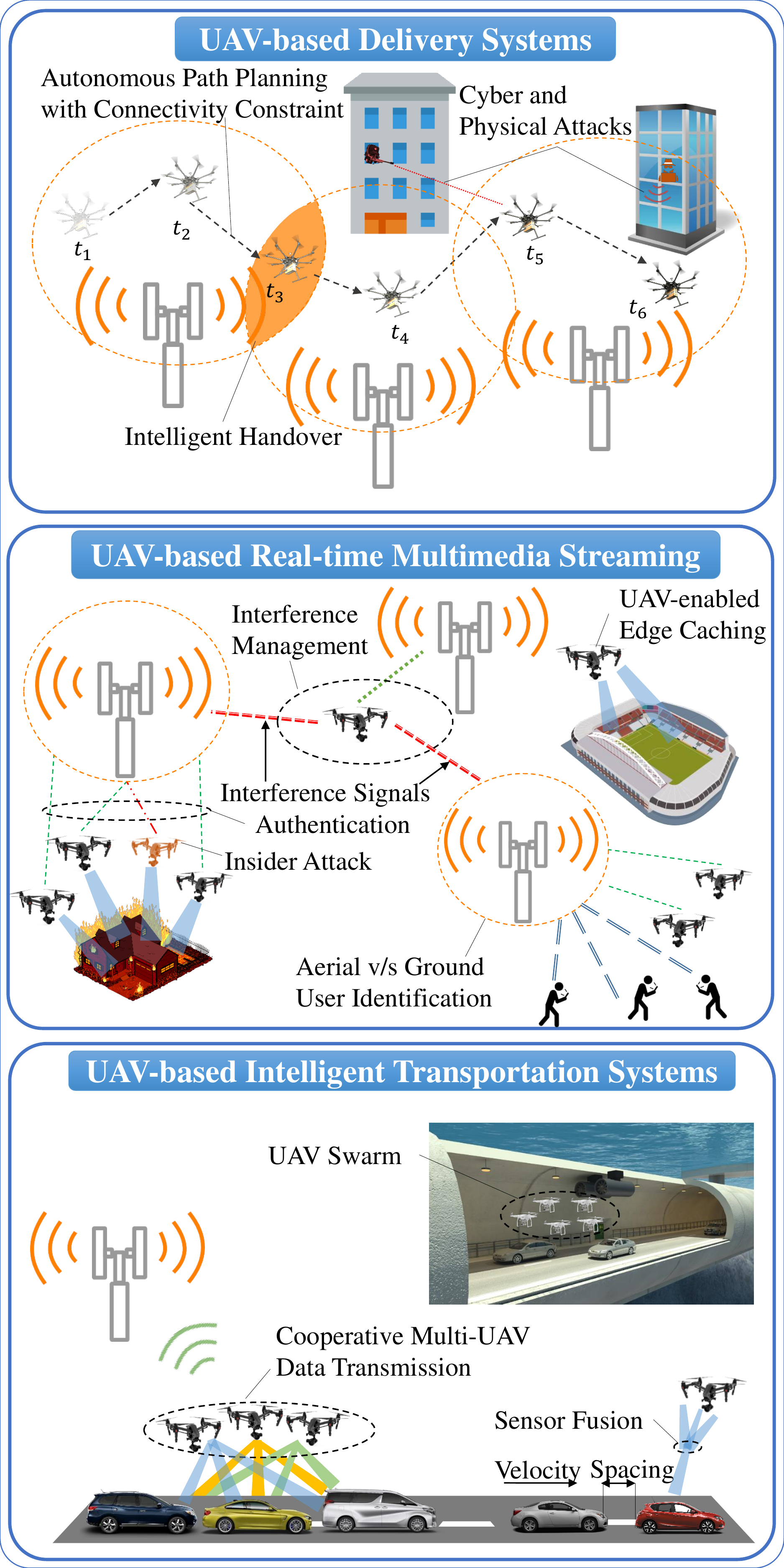}
   \caption{Examples of wireless and security challenges of cellular-connected UAVs in UAV-based delivery systems, UAV-based real-time multimedia streaming networks, and UAV-enabled intelligent transportation systems.}\label{UAV_challenges}
  \end{center}
  \vspace{-0.6cm}
\end{figure}

The main contribution of this paper is to expose the major wireless and security challenges that arise in different UAV-based applications and suggest artificial neural network (ANN) based solution approaches for addressing such challenges. In particular, we focus on three major use cases for cellular-connected UAVs: UAV-based delivery systems, UAV-based real-time multimedia streaming networks, and UAV-enabled intelligent transportation systems. For each one of these use cases, we introduce the main technical challenges, in terms of wireless connectivity and security (as illustrated in Figure~\ref{UAV_challenges}), while outlining new AI-inspired solutions to address those challenges. The introduced AI solutions enable the UAVs to predict future network changes thus adaptively optimizing their actions in order to efficiently manage their resources while securing a safe operation. We also provide preliminary simulation results to showcase the benefits of the introduced solutions for each cellular-connected UAV application use case. Here, we restrict our attention to the security at higher communication layers since physical layer security issues and solutions have been discussed in \cite{security_ref}.

The rest of this paper is organized as follows. Section~\ref{section:UAV_DS} presents the communication and wireless challenges in UAV-DS and proposes AI-based solution schemes for such challenges. Section~\ref{section:UAV_RMS} highlights the main communication and security challenges in UAV-RMS applications and the corresponding proposed ANN-based solutions. Section~\ref{section:UAV_ITS} provides ANN-based solution schemes for the main communication and security challenges in UAV-ITSs. Finally, conclusions are given in Section~\ref{section:conc}.


\section{UAV-Based Delivery Systems}\label{section:UAV_DS}

\subsection{Motivation}
UAV-based delivery systems have received much attention recently for various applications such as postal and package delivery (e.g., Amazon prime), food delivery, transport of medicines and vaccinations, and drone taxis for delivery of people~\cite{UAV_delivery}. Compared to conventional delivery methods, UAV-DSs allow a faster delivery process at a reduced cost. They can also provide mission critical services reaching remote and inaccessible areas. To reap the benefits of UAV-DSs, it is important to provide cellular connectivity to the UAVs for control and signaling data transmission. In essence, providing cellular connectivity to delivery UAVs allows network operators to track their location and guarantee a secure delivery of the transported goods. Therefore, to realize such benefits, it is important to address several wireless and security challenges related to cellular-connected UAV-DSs, ranging from efficient handover and path planning to cyber-physical attacks.


%
%
%
%



\subsection{Wireless Challenges and AI Solutions}

\subsubsection{Ultra-Reliable and Low-Latency Communications (URLLC)}\label{URLLC}
In UAV-DS, the UAVs must send critical control information while delivering goods to their destinations. This, in essence, requires latency of $1$ ms or less and exceedingly stringent reliability with a target block error rate as low as $10^{-5}$~\cite{URLLC_requirements}, especially in mission-critical scenarios such as medical delivery. Wireless latency encompasses both signaling overhead and data transmission. To achieve low signaling latency, channel estimation can be predicted in advance using AI thus allowing a proactive allocation of radio resources. This can be realized by incorporating a long-short term memory (LSTM) cell at the UAV level for learning a sequence of future channel states~\cite{ursula_TWC_1}. LSTMs are effective in dealing with long term dependencies which makes them suitable for learning a sequence of a time-dependent vector. 
Moreover, in a large network of UAVs, constantly communicating with a remote cloud can introduce substantial communication and signaling delays. To reduce such delays, one can rely on on-device machine learning or edge AI. As opposed to centralized, cloud-based AI schemes, on-device machine learning is based on a distributed machine learning approach, such as \emph{federated learning} (FL), in which the training data describing a particular AI task (e.g., resource management or computing) is stored in a distributed fashion across the UAVs and the optimization problem is solved collectively~\cite{federated_learning}. This in turn enables a large number of UAVs to collaboratively allocate their radio resources in a distributed way thus reducing wireless congestion and device-to-cloud latency. Finally, it is important to note that transmission latency can be further reduced by improving wireless connectivity, as discussed in Section~\ref{multimedia_wireless}.







\subsubsection{Efficient Handover}
In UAV-DSs, the UAVs face frequent handovers and handover to distant cells thus resulting in a ping-pong effect. As opposed to ground UEs, cellular-connected UAVs exhibit LoS links with multiple neighboring BSs simultaneously which, along with dynamic channel variations, can result in a fluctuation in the quality of their wireless transmission. In this context, it is necessary to have complete and sequential information about the channel signal quality at different locations before and after the current location of a particular UAV. As such, bidirectional LSTM cells (bi-LSTMs) are suited for addressing this challenge as they exploit both the previous and the future context, by processing the input data (i.e., channel quality) from two directions with two separate hidden layers. In particular, one LSTM layer processes the input sequence in the forward direction, while the other LSTM layer processes the input in the reverse direction~\cite{bi-LSTM}. Therefore, instead of accounting for the next time step only, this scheme enables each UAV to consider the channel quality at its previous and future sequence locations. This framework can hence be trained to allow the UAVs to update their corresponding cell association vector while avoiding frequent handovers based on previous and future channel signal quality.


\subsubsection{Autonomous Path Planning with Connectivity Constraints}
A critical factor for UAV-DSs is to maintain reliable cellular connectivity for the UAVs at each time instant along their corresponding paths while also minimizing the total time required to accomplish their delivery mission. In essence, a delivery UAV must maintain a minimum signal-to-noise-and-interference (SINR) ratio along its path to guarantee a reliable communication link for its control information. This naturally depends on the UAV's location, cell association vector, transmit power level, and the location of the serving ground BS.
As such, a key challenge for UAV-DSs is to optimize the UAVs' paths so as to reduce their total delivery time while guaranteeing reliable wireless connectivity and thus an instantaneous SINR threshold value. Although a centralized approach can update the path plan of each UAV, this would require real-time tracking of the UAVs and control signals to be transmitted to the UAVs at all time. Moreover, a centralized approach incurs high round-trip latencies and requires a central entity to acquire full knowledge of the current network state. To overcome these challenges, \emph{online} edge algorithms must be implemented individually by each UAV to plan its future path. In this regard, convolutional neural networks (CNNs) can be combined with a deep reinforcement learning (RL) algorithm based on a recurrent neural network (RNN) (e.g., echo state network (ESN) or LSTM) at the UAV level resulting in a CNN-RNN scheme. ESN exhibits dynamic temporal behavior and is characterized by its adaptive memory that enables it to store necessary previous state information to predict the future steps of each UAV. Meanwhile, CNN are mainly used for image recognition and thus can be used for identifying the UAV's environment by extracting features from input images. For instance, CNNs aid the UAVs in identifying the location of the ground BSs, ground UEs, and other UAVs in the network. These extracted features are then fed to a deep RNN which can be trained to learn an optimized sequence of the UAV's future steps, that would minimize its delivery time and guarantee a reliable cellular connectivity at each time instant, based on the input features.

\begin{figure}[t!]
  \begin{center}
  \centering
   \includegraphics[width=9cm]{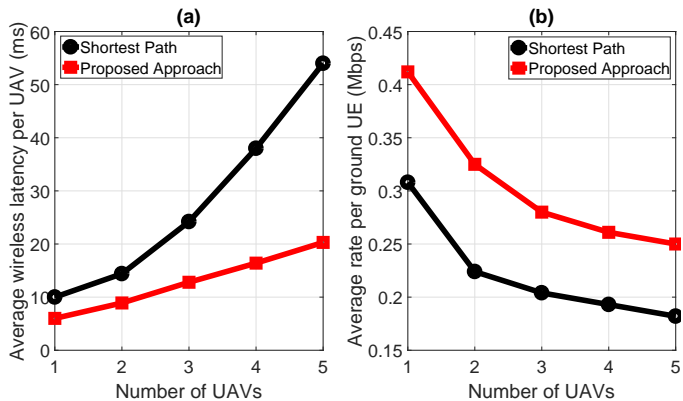}
   \caption{Performance assessment of the proposed deep ESN-based path planning algorithm in terms of average (a) wireless latency per UAV and (b) rate per ground UE as compared to the shortest path approach, for different number of UAVs~\cite{ursula_path_planning}.}\label{path_planning}
  \end{center}
  \vspace{-0.6cm}
\end{figure}

In this regard, in~\cite{ursula_path_planning}, we proposed a deep RL framework based on ESN (D-ESN) for optimizing the trajectories of multiple cellular-connected UAVs in an online manner while minimizing latency and interference. For simplicity, we consider an input vector describing the locations of the neighboring ground BSs and other UAVs instead of extracting such features from a CNN. To highlight the gain of D-ESN for path planning, we compare the average values of the (a) wireless latency per UAV and (b) rate per ground UE resulting from the proposed path planning scheme and the shortest path scheme, as shown in Fig.~\ref{path_planning}. Clearly, from Fig.~\ref{path_planning}, we can see that exploiting a deep ESN-based path planning scheme under connectivity constraints for cellular-connected UAVs results in a more reliable wireless connectivity and in lower latency, as compared to wireless-unaware shortest path scheme.

\subsection{Security Challenges and AI Solutions}
Due to the UAVs' altitude limitations and the LoS communication link with the ground BS, UAV-based delivery systems are vulnerable to \emph{cyber-physical (CP) attacks} in which an adversary aims at compromising a delivery UAV, taking over its control, and ultimately destroying, delaying, or stealing the transported goods. To thwart such CP attacks, the UAV can create a CP threat map in which the adversaries locations can be categorized based on the environmental objects where the UAVs can be physically attacked as well as communication network where the cyber attacks can be imposed to the communication link. Even though prior works assume that a threat map is predetermined\cite{Sanjab}, it is important to create such map in an online manner in order to account for real-time changes in the environment and to overcome the memory limitation of the UAVs for storing a large-scale map. To realize this, a CNN can be trained for classifying the high-risk locations by taking as input the images of the UAV's surrounding environment along each position of its path. 
From the operator's perspective, it is also important to detect any potential attack by identifying any abnormal or undesirable behavior in the UAVs' motion. Therefore, given their capability of dealing with time-series data, RNNs can be adopted for capturing the UAV's motion characteristics by feeding them with the UAV's dynamics such as its position, speed, acceleration, and destination location. In this case, the RNN's output will be the predicted UAV's normal motion and, thus, using this output the operator can distinguish UAV's abnormal motion which is resulted from a CP attack.

\section{UAV-Based Real-Time Multimedia Streaming Applications}\label{section:UAV_RMS}

\subsection{Motivation}
One key use case for cellular-connected UAVs is to provide various real-time multimedia streaming applications such as online video streaming and broadcasting, UAV-enabled virtual reality (VR), online tracking and localization of mobile targets, and surveillance. In essence, providing cellular connectivity to the UAVs enables online transmission of data and low-latency wireless communication which are essential factors for multimedia streaming applications.
To enable effective delivery of such real-time multimedia using cellular-connected UAVs, several wireless and security challenges need to be addressed ranging from interference management to authentication.

\subsection{Wireless Challenges and AI Solutions}\label{multimedia_wireless}
\subsubsection{Interference Management}
For UAV-RMS applications, UAVs will mainly transmit data in the \emph{uplink}. Nevertheless, the ability of cellular-connected UAVs to establish LoS connectivity with multiple ground BSs can lead to substantial mutual interference among them as well as to the ground users. To address this challenge, new improvements in the design of future cellular networks such as advanced receivers, cell coordination, 3D frequency reuse, and 3D beamforming, are needed. For instance, due to their ability of recognizing and classifying images, CNNs can be implemented on each UAV in order to identify several features of the environment such as the location of UAVs, BSs, and ground UEs. Such an approach will enable each UAV to adjust its beamwidth tilt angle so as to minimize the interference on the ground UEs. Moreover, in streaming scenarios, UAV trajectory optimization is also essential. In particular, physical layer solutions such as 3D beamforming, can be combined with an interference-aware path planning scheme to guarantee more efficient communication links for both ground and aerial users. Such a path planning scheme (e.g., such as the one we proposed in~\cite{ursula_path_planning}) allows the UAVs to adapt their movement based on the rate requirements of both aerial UAV-UEs and ground UEs, thus improving the overall network performance.



\subsubsection{UAV-enabled Edge Caching}
For various real-time multimedia streaming applications, cellular-connected UAVs must generate videos from data files collected using sensors and cameras. For instance, in UAV-enabled VR applications, the UAVs will generate $360^\circ$ videos for each user. However, each UAV can only collect a limited number of data files which might not be sufficient for generating all the requested videos. Meanwhile, cache-enabled UAVs can store common data files related to popular content or for generating videos that users may request in the future thus reducing the number of data files that UAVs need to collect when a request is made~\cite{mingzhe_caching}. For instance, for UAV-enabled VR applications, cache-enabled UAVs can directly store a $360^\circ$ video and send a rotated version of this stored video according to each user's viewing perspective. Moreover, for game broadcast applications, cache-enabled UAVs can store the environment of the game and thus would only need to track the motions of the players for updating the cached data. Here, CNNs can once again be adopted for allowing cache-enabled UAVs to store popular videos or common data files. In particular, CNNs can extract and store the common features of the data files that are requested by different users or by each user, at different time slots. Furthermore, CNNs can be used to record the features of each UAV's surrounding environment. Consequently, when the UAVs need to collect data in a new environment, they would only need to collect new features that are not already recorded by the CNNs. In this context, RNNs can also be employed for predicting the users' video requests. In fact, the context requests of users can be correlated over time, and thus, RNNs can enable the UAVs to cache in advance the predicted future requests or other popular multimedia files.

\begin{figure}[t]
\centering
\subfigure[]{{\setlength{\belowdisplayskip}{10 pt}}
\label{figure2a} 
\includegraphics[width=6.5cm]{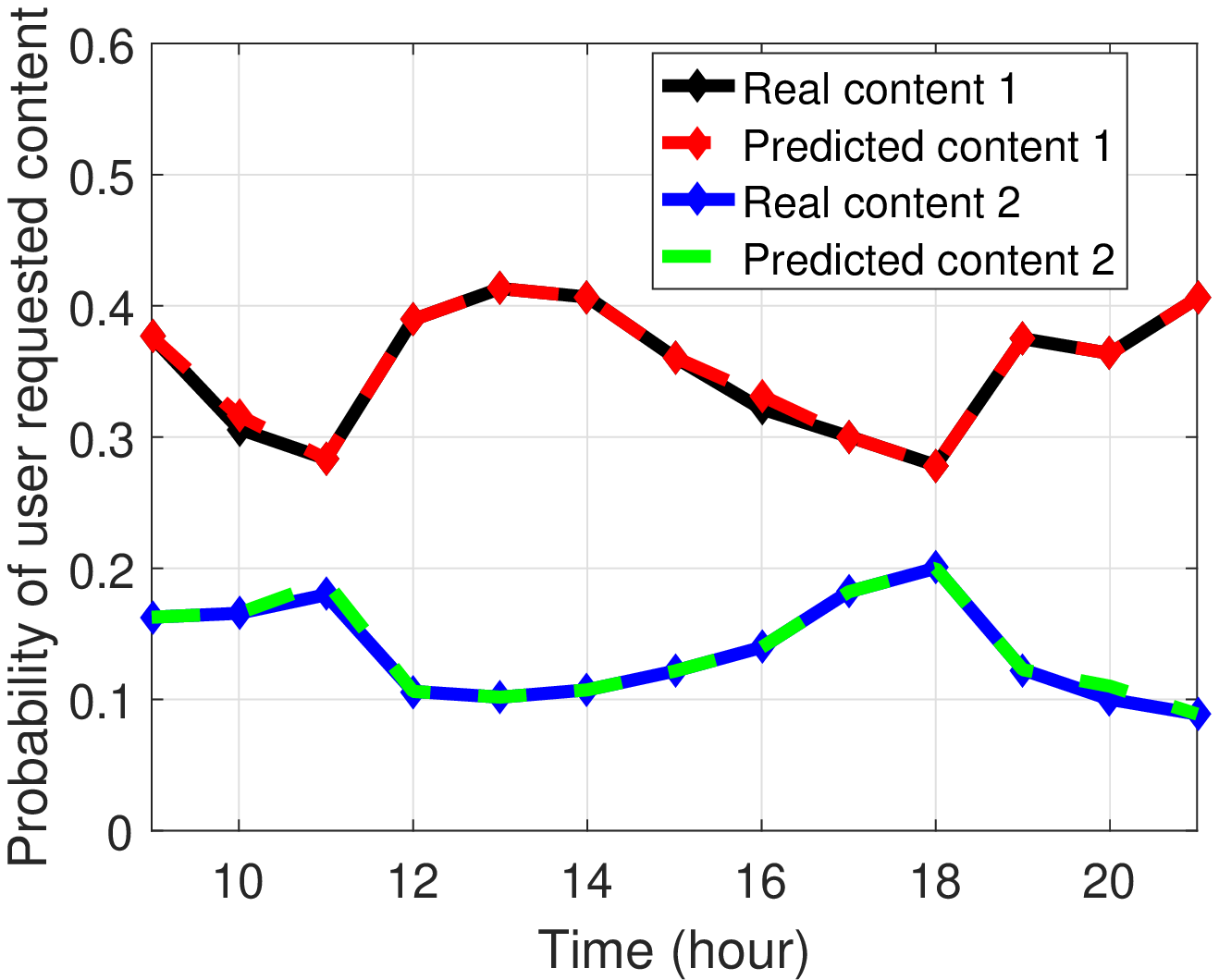}}
\subfigure[]{
\label{figure2b} 
\includegraphics[width=6cm]{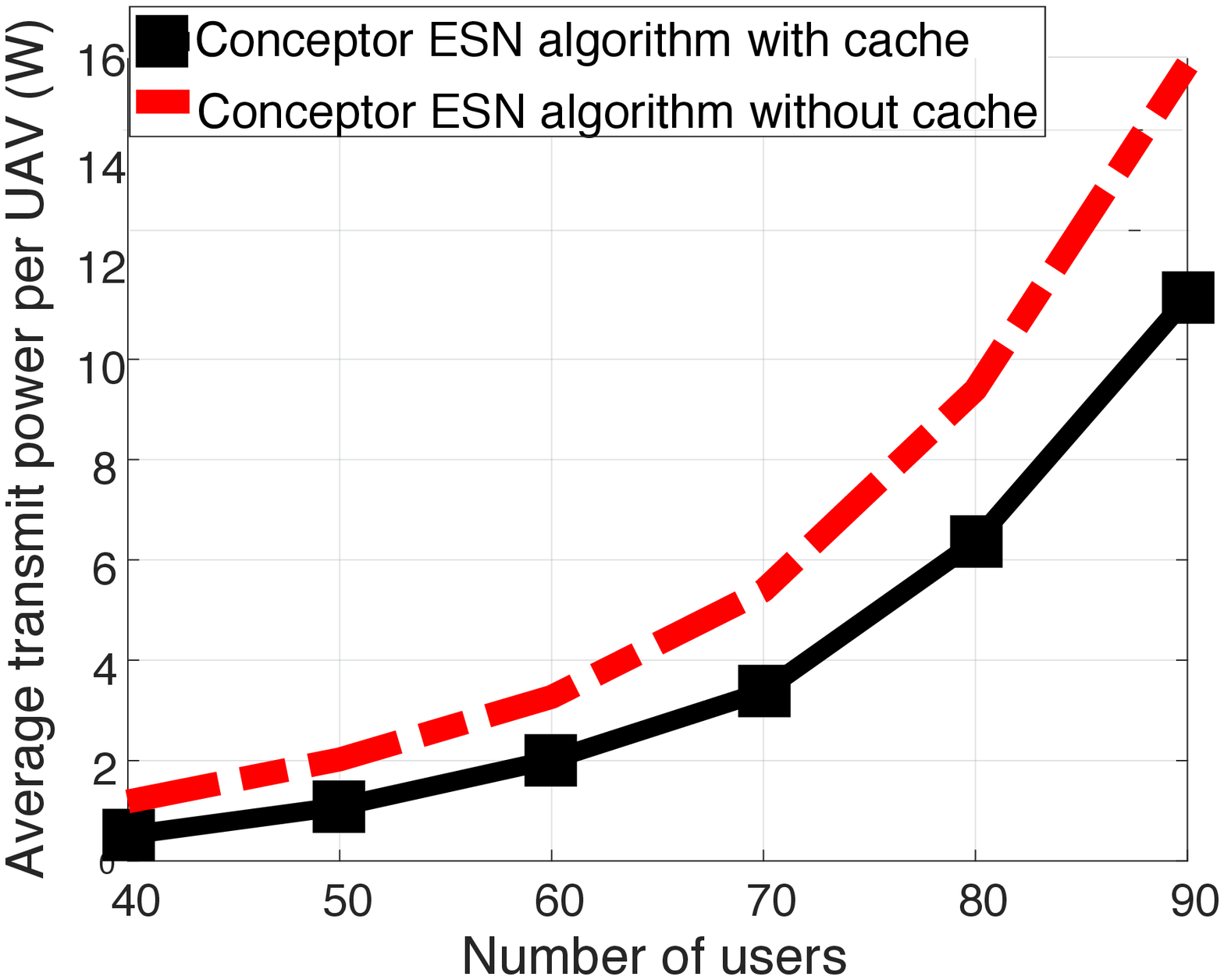}}
  \vspace{-0.4cm}
 \caption{\label{figure2b} (a) Comparison of the content request probability predictions for the proposed conceptor ESN algorithm with the real data and (b) the average UAV transmit power as a function of the number of users in the network for the proposed conceptor ESN algorithm with and without caching~\cite{mingzhe_caching}.}
  \vspace{-0.75cm}
\end{figure}

Based on our work in~\cite{mingzhe_caching}, we introduce an ESN-based algorithm for predicting the user's content request distributions. The input to the proposed framework is the users' context information such as age, gender, and job, and the output of the ESN-based algorithm is the distribution of the users' content requests. Therefore, based on the users' content request distributions, the UAVs can determine the contents to store at the UAV cache and, thus, transmit the cached contents to the users without the need for backhaul connections. Using real data from \emph{Youku}, Fig. \ref{figure2a} shows that the ESN-based algorithm can accurately predict the content request distribution of a given user. Fig. \ref{figure2b} (b) shows the average transmit power per UAV of cache-enabled UAVs as a function of the number of users. In Fig. \ref{figure2b}, we can see that the proposed ESN algorithm for cache-enabled UAVs yields a considerable reduction in transmit power compared to a baseline without caching.  


\subsubsection{Identification of Aerial and Ground Users}
As shown in~\cite{sebastien}, the radio propagation environment experienced by cellular-connected UAVs differs from that experienced by ground users. Consequently, to maximize the total network performance, a network operator must allocate its radio resources differently between airborne and ground users, especially for UAV-RMS applications. To realize this, network operators should be capable of differentiating an airborne user from a ground one which cannot be achieved by solely relying on self-reporting due to the possibility of a faulty report. Instead, network operators can utilize wireless cellular radio measurements such as reference signal received power (RSRP), received signal strength indicator (RSSI), and reference signal received quality (RSRQ) for user classification. These features can essentially act as an input to a deep belief network (DBN) which can be trained for classifying an airborne user from a ground one. In essence, DBNs are deep architectures that consist of a stack of restricted Boltzmann machines (RBMs), thus having the benefit that each layer can learn more complex features than the layers before it. In essence, a pre-training step is done in DBNs thus overcoming the vanishing gradient problem. This is then followed by fine-tuning the network weights using conventional error back propagation algorithm. 

\vspace{-0.1cm}
\subsection{Security Challenges and AI Solutions}


In UAV-RMS applications, an attacker can disrupt the UAV's data transmissions by forging the identities of the transmitting UAVs' and sending disrupted data using their identity. This type of \emph{insider attacks} becomes particularly acute in a large-scale UAV system. In particular, the BS must process the received multimedia files from all the UAVs and allocate computational resources for authenticating the UAVs. However, in large-scale networks, authenticating all the UAVs at once exceeds the BS's computational resources, thus, incurring delay for processing the received files.
To avoid this delay, the BS can authenticate only a fraction of the UAVs at each time step. To realize this, the BS could implement a deep RL algorithm based on LSTM in order to learn what signals to authenticate at each time step of its authentication process. In particular, this framework takes as an input a sequence of previous security states of each UAV indicating whether a UAV was previously vulnerable to attacks, and learns a sequence of future authentication decisions for each UAV. LSTMs are suitable for this application since they can learn the interdependence of UAVs' vulnerability at the past time steps, memorize the importance of UAVs to the BS, and map the past sequence of UAV states to a future decision sequence.



\begin{figure}[t!]
	\begin{center}
		\centering
		\includegraphics[width=8cm]{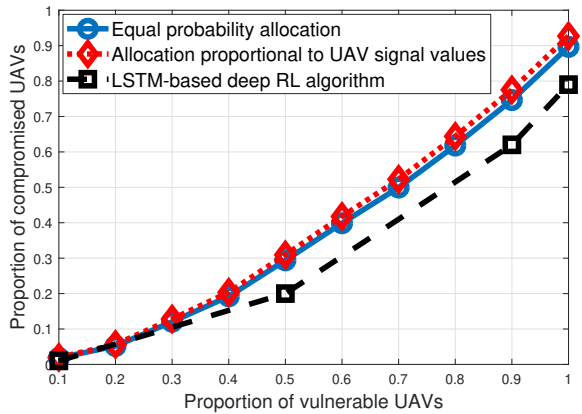}
		\caption{The proportion of compromised cellular-connected UAVs as a function of the proportion of vulnerable UAVs in a large-scale UAV system authentication~\cite{ferdowsi2018deep}.}\label{fig:authentication}
	\end{center}
	\vspace{-0.6cm}
\end{figure}

To analyze the performance of LSTM-based deep RL method for authentication, based on \cite{ferdowsi2018deep}, we consider a network of 1000 UAVs that transmit multimedia streams to a BS. We analyze different scenarios in which different proportions of available UAVs are vulnerable to cyber attacks. 
Fig. \ref{fig:authentication} assesses the performance of the LSTM-based deep RL framework compared to two baseline authentication scenarios. From Fig. \ref{fig:authentication}, we can see that the proposed algorithm performs the same as the two other baselines in the low range of proportion of vulnerable UAVs. However, as the number of vulnerable UAVs increases the LSTM-based deep RL outperforms the two other baselines and reduces the proportion of compromised UAVs in the network.

\section{UAV-Enabled Intelligent Transportation Systems}\label{section:UAV_ITS}
\vspace{-0.2cm}
\subsection{Motivation}
Integrating UAVs in an intelligent transportation system (ITS) would control road traffic, monitor incidents, and enforce road safety. For instance, UAVs can provide a quick report in case of an accident and can act as flying roadside units, speed cameras, and dynamic traffic signals. Moreover, for vehicular platoons, to reduce wireless network congestion, a cellular-connected UAV can send control and network related information to one of the vehicles only and this vehicle can share the information with other vehicles in the platoon via dedicated short range communication links. UAVs can also track the behavior of a platoon thus detecting any compromised vehicle.
Therefore, to reap the benefits of UAV-ITS, several wireless and security challenges need to be addressed ranging from cooperative multi-UAV data transmission and multimodal data integration to secured consensus of UAV swarms. 

\subsection{Wireless Challenges and AI Solutions}

\begin{table*}[t!]\footnotesize
\setlength{\belowcaptionskip}{0pt}
\setlength{\abovedisplayskip}{3pt}
\captionsetup{belowskip=0pt}
\newcommand{\tabincell}[2]{\begin{tabular}{@{}#1@{}}#1.6\end{tabular}}
 \setlength{\abovecaptionskip}{2pt}
 \renewcommand{\captionlabelfont}{\small}
 \captionsetup{justification=centering}
\caption{Cellular-connected UAV use cases, challenges, and ANN-based solution schemes.}\label{UAV_table}
\centering
\tabcolsep=0.03cm 
\begin{tabular}{|c|c|c|c||c|c|c|c|c|c|c|c|c|c|c|c|c|}
\hline
\textbf{Wireless and Security Challenges}& \multicolumn{3}{|c||}{\textbf{UAV-based Applications}} & \multicolumn{10}{|c|}{\textbf{ANN-based Solutions}}\\
\hline
& UAV-DS & UAV-RMS & UAV-ITS & FL & bi-LSTM & CNN-RNN & D-ESN & CNN & ESN & DBN & LSTM & DSC & m-RBM\\
\hline
URLLC &\checkmark & & &\checkmark & & & & & & & & &\\
\hline
Efficient Handover &\checkmark & & & &\checkmark & & & & & & & &\\
\hline
Autonomous Path Planning  &\checkmark & & & & &\checkmark &\checkmark & & & & & & \\
\hline
Interference Management & &\checkmark & & & & & &\checkmark & & & & & \\
\hline
UAV-enabled Edge Caching & &\checkmark & & & & & &\checkmark &\checkmark & & & & \\
\hline
Identification of Aerial and Ground Users & &\checkmark & & & & & & & & \checkmark& & & \\
\hline
Cooperative Multi-UAV Data Transmission & & &\checkmark & & & & & & & & &\checkmark & \\
\hline
Multimodal Sensor Fusion & & &\checkmark & & & & & & & & & &\checkmark \\
\hline
Cyber-Physical Attacks &\checkmark & & & & & & &\checkmark & & & & & \\
\hline
Authentication of UAVs & &\checkmark & & & & & & & & & \checkmark & & \\
\hline
Secured Consensus of UAV Swarms & & &\checkmark &\checkmark & & & & & & & & & \\
\hline
\end{tabular}
\vspace{-0.24cm}
\end{table*}

\subsubsection{Cooperative Multi-UAV Data Transmission}
In UAV-ITSs, each UAV is generally equipped with multiple sensors such as LiDAR and GPS and would therefore need to send different types of multimedia files and/or big data (e.g., 3D-map representation of the environment) to either other UAVs, vehicles, or the infrastructure, simultaneously. 
In such scenarios, it would be essential for different UAVs in a given geographical area to coordinate their data transmission. In other words, instead of each UAV transmitting the whole data file, e.g., area map, to its corresponding vehicle, each UAV will transmit a different part of the data file to all of the vehicles in a given geographical area thus resulting in a faster data transmission and a lower power consumption per UAV. In this regard, deep spectral clustering (DSC) learning can be adopted for grouping the UAVs into several clusters for data transmission based on their location, type of sensors they encompass, data files they need to transmit, and the location and number of vehicles in the network. In essence, DSC learns a map that embeds this input data into the eigenspace of their associated graph Laplacian matrix and thus clusters them accordingly. Consequently, DSC endows the UAVs with the capability of transmitting correlated data in a cooperative and distributed manner to the vehicles. This would essentially result in a faster data transmission to the vehicles thus allowing them to make real-time decisions for a safe navigation among the surrounding traffic. DSC can be combined with cooperative game theory, for further analysis of cooperative swarms of UAVs. Moreover, the presence of high mobility in ITSs along with cooperative UAV swarms, requires revisiting the interference and resource management schemes of Sections~\ref{section:UAV_DS} and~\ref{section:UAV_RMS} to handle the more dynamic and cooperative ITS environment.

\subsubsection{Multimodal Sensor Fusion}
In UAV-ITS, UAVs must transmit each one of their sensor readings to other network nodes, thus, resulting in cellular network congestion in case of dense UAV deployment. However, energy consumption and bandwidth allocation are important factors that determine the maximum operation time of the UAVs. As such, to reduce the power and bandwidth allocated for transmitting the sensor readings, a UAV can integrate its heterogeneous sensor readings into one vector thus resulting in less data transmissions over the UAV-vehicle links while also providing a more comprehensive assessment of the environment. Nevertheless, there exists differences between sensors ranging from sampling rates to the data generation model thus making UAV-based ITS sensor integration challenging. In this regard, multimodal RBMs (m-RBMs) are a suitable tool for combining different perspectives captured in signals of multimodal data for a system with multiple sensors~\cite{multimodal}. A m-RBM can be implemented at the UAV level thus identifying nonintuitive features largely from cross-sensor correlations which can yield accurate estimation. From the UAV's perspective, this approach enables each UAV to have a better assessment of its environment. For instance, a system trained simultaneously to detect an accident, high speed vehicle, and an anomalous vehicle does better than three separate systems trained in isolation since the single network can share information among the separate tasks. From the wireless network perspective, multimodal sensor fusion improves the UAV's energy efficiency and results in less data transmissions over the UAV-vehicle links thus reducing wireless congestion and enabling a larger number of UAVs to be served simultaneously.


\subsection{Security Challenges and AI Solutions}
For UAV-ITS, a swarm of coordinated UAVs has the capability of performing missions compared to single UAVs. Swarming UAVs communicate with each other while in flight to reach a consensus over their defined task, and can respond to changing conditions autonomously. A good analogy would be a dense flock of starlings reacting to a sudden threat like a hawk. Nevertheless, this data sharing scheme among a swarm of UAVs is generally prone to \emph{adversarial machine learning} attacks in which an attacker can join the swarm and alter their shared data, which results in non-harmonious movements as well as collisions. To overcome this challenge, federated learning can be adopted for a swarm of UAVs. In federated learning, each UAV receives the common task that needs to be accomplished by the UAV swarm from the BS and improves its learning model for completing the required tasks based on its collected data only. Then, each UAV summarizes the changes in its learning model and shares this summary with other UAVs in the swarm. This, indeed, will solve the vulnerability of raw data transmission between the UAVs and thus mitigating the risk of the adversarial machine learning.

Table~\ref{UAV_table} provides a summary of the wireless and security challenges of cellular-connected UAVs in UAV-DS, UAV-RMS, and UAV-ITS while suggesting ANN-based solution schemes.

\section{Conclusion}\label{section:conc}
In this paper, we have summarized the main use cases of cellular-connected UAVs in UAV-DS, UAV-RMS, and UAV-ITS applications. We have highlighted the main wireless and security challenges that arise in such scenarios while introducing various AI-based solutions for addressing such challenges. Preliminary simulation results have shown the benefits of the introduced solutions for each cellular-connected UAV application use case.

\bibliographystyle{IEEEtran}
\bibliography{references}

\begin{thebibliography}{10}
\providecommand{\url}[1]{#1}
\csname url@samestyle\endcsname
\providecommand{\newblock}{\relax}
\providecommand{\bibinfo}[2]{#2}
\providecommand{\BIBentrySTDinterwordspacing}{\spaceskip=0pt\relax}
\providecommand{\BIBentryALTinterwordstretchfactor}{4}
\providecommand{\BIBentryALTinterwordspacing}{\spaceskip=\fontdimen2\font plus
\BIBentryALTinterwordstretchfactor\fontdimen3\font minus
  \fontdimen4\font\relax}
\providecommand{\BIBforeignlanguage}[2]{{%
\expandafter\ifx\csname l@#1\endcsname\relax
\typeout{** WARNING: IEEEtran.bst: No hyphenation pattern has been}%
\typeout{** loaded for the language `#1'. Using the pattern for}%
\typeout{** the default language instead.}%
\else
\language=\csname l@#1\endcsname
\fi
#2}}
\providecommand{\BIBdecl}{\relax}
\BIBdecl

\bibitem{tutorial_ML}
M.~Chen, U.~Challita, W.~Saad, C.~Yin, and M.~Debbah, ``Machine learning for
  wireless networks with artificial intelligence: A tutorial on neural
  networks,'' \emph{arXiv:1710.02913}, Oct. 2017.

\bibitem{sebastien}
S.~Euler, H.~Maattanen, X.~Lin, Z.~Zou, M.~Bergstr\"{o}m, and J.~Sedin,
  ``Mobility support for cellular connected unmanned aerial vehicles:
  Performance and analysis,'' \emph{arXiv:1804.04523}, Apr. 2018.

\bibitem{mozaffari_survey}
M.~Mozaffari, W.~Saad, M.~Bennis, {Y. H. Nam}, and M.~Debbah, ``A tutorial on
  {UAVs} for wireless networks: Applications, challenges, and open problems,''
  \emph{arXiv:1803.00680}, Mar. 2018.

\bibitem{ismail}
E.~Bulut and I.~G$\ddot{\mathrm{u}}$ven\c{c}, ``Trajectory optimization for
  cellular-connected {UAVs} with disconnectivity constraint,'' in \emph{Proc.
  of International Conference on Communications (ICC)- Integrating UAVs into 5G
  (UAV-5G)}.\hskip 1em plus 0.5em minus 0.4em\relax Kansas City, MO, USA, May
  2018.

\bibitem{security_ref}
G.~Zhang, Q.~Wu, M.~Cui, and R.~Zhang, ``Securing {UAV} communications via
  joint trajectory and power control,'' \emph{arXiv:1801.06682}, Jan. 2018.

\bibitem{UAV_delivery}
P.~Grippa, D.~Behrens, C.~Bettstetter, and F.~Wall, ``Job selection in a
  network of autonomous {UAVs} for delivery of goods,'' in \emph{Proc. of
  Robotics: Science and Systems (RSS)}.\hskip 1em plus 0.5em minus 0.4em\relax
  Cambridge, Massachusetts, USA, July 2014.

\bibitem{URLLC_requirements}
J.~Nielsen, R.~Liu, and P.~Popovski, ``Ultra-reliable low latency communication
  using interface diversity,'' \emph{IEEE Transactions on Communications},
  vol.~66, no.~3, pp. 1322--1334, Nov. 2017.

\bibitem{ursula_TWC_1}
{U. Challita, and L. Dong, and W. Saad}, ``Proactive resource management in
  {LTE-U} systems: A deep learning perspective,'' \emph{arXiv:1702.07031}, Feb.
  2017.

\bibitem{federated_learning}
J.~Konecn$\acute{\mathrm{y}}$, H.~B. McMahan, D.~Ramage, and
  P.~Richt$\acute{\mathrm{a}}$rik, ``Federated optimization: Distributed
  machine learning for on-device intelligence,'' \emph{arXiv:1610.02527}, Oct.
  2016.

\bibitem{bi-LSTM}
A.~Zeyer, P.~Doetsch, P.~Voigtlaender, R.~Schl$\ddot{\textrm{u}}$ter, and
  H.~Ney, ``A comprehensive study of deep bidirectional {LSTM RNNs} for
  acoustic modeling in speech recognition,'' \emph{arXiv:1606.06871}, Mar.
  2017.

\bibitem{ursula_path_planning}
{U. Challita, and W. Saad, and C. Bettstetter}, ``Cellular-connected {UAVs}
  over {5G}: Deep reinforcement learning for interference management,''
  \emph{arXiv:1801.05500}, Jan. 2018.

\bibitem{Sanjab}
A.~Sanjab, W.~Saad, and T.~Ba\c{s}ar, ``Prospect theory for enhanced
  cyber-physical security of drone delivery systems: A network interdiction
  game,'' in \emph{Proc. of IEEE International Conference on Communications
  (ICC)}, Paris, France, 2017.

\bibitem{mingzhe_caching}
M.~Chen, M.~Mozaffari, W.~Saad, C.~Yin, M.~Debbah, and {C. S. Hong}, ``Caching
  in the sky: Proactive deployment of cache-enabled unmanned aerial vehicles
  for optimized quality-of-experience,'' \emph{IEEE Journal on Selected Areas
  in Communications}, vol.~35, no.~5, pp. 1046--1061, May 2017.

\bibitem{ferdowsi2018deep}
A.~Ferdowsi and W.~Saad, ``Deep learning for signal authentication and security
  in massive {Internet of Things} systems,'' \emph{arXiv preprint
  arXiv:1803.00916}, 2018.

\bibitem{multimodal}
N.~Srivastava and R.~Salakhutdinov, ``Multimodal learning with deep boltzmann
  machines,'' in \emph{Proc. of Advances in Neural Information Processing
  Systems (NIPS)}.\hskip 1em plus 0.5em minus 0.4em\relax Lake Tahoe, CA, USA,
  Dec. 2012.

\end{thebibliography}

\end{document}